\documentclass[a4paper,fleqn,usenatbib,backref]{mnras}
\usepackage{longtable}
\usepackage{amsmath}
\usepackage{enumerate}
\usepackage{graphicx}
\usepackage{diagbox}
\usepackage{url}
\usepackage{float}
\usepackage{subfigure}
\usepackage{ulem}       %strikeout, added by author
\usepackage{color,xcolor} % textcolor, added by author
\usepackage{makecell}

\title[Giant pulses from PSR J1047$-$6709]{Detection of giant pulses in PSR J1047$-$6709}
\author[S. N. Sun et al.]{S. N. Sun$^{1,2}$, W. M. Yan$^{1,3,4}$\thanks{E-mail: yanwm@xao.ac.cn}, N. Wang$^{1,3,4}$
\\
$^{1}$Xinjiang Astronomical Observatory, Chinese Academy of Sciences, Urumqi, Xinjiang 830011, China\\
$^{2}$University of Chinese Academy of Sciences, Beijing 100049, China\\
$^{3}$Key Laboratory of Radio Astronomy, Chinese Academy of Sciences, Nanjing 210008, China\\
$^{4}$Xinjiang Key Laboratory of Radio Astrophysics, 150 Science1-Street, Urumqi, Xinjiang, 830011, People's Republic of China\\
\
\
\
}

\date{Accepted XXX. Received YYY; in original form ZZZ}

\pubyear{2020}

\begin{document}
\label{firstpage}
\pagerange{\pageref{firstpage}--\pageref{lastpage}}
\maketitle

\begin{abstract}
We report the emission variations in PSR J1047$-$6709 observed at 1369 MHz using
the Parkes 64 m radio telescope. This pulsar shows two distinct emission states:
a weak state and a bright emission state. We detected giant pulses (GPs)
in the bright state for the first time. We found 75 GPs with pulse width ranging from  0.6 to 2.6 ms. The energy of GPs follows a power-law distribution with the index $\alpha=-3.26\pm0.22$. The peak flux density of the brightest GP is 19 Jy which is 110 times stronger than the mean pulse profile.
The polarization properties of the average profile of GPs are similar to
that of the pulses with energy less than 10 times average pulse energy in the bright state. This indicates that the emission mechanism is basically 
the same for them. Our results provide a new insight into the origin of the GPs in pulsars.

\end{abstract}

\begin{keywords}
{stars: neutron - pulsars: general - pulsars: individual (PSR J1047$-$6709)}
\end{keywords}

\section{Introduction}\label{sec:intro}

A giant pulse (GP) is the short-duration, burst-like radio
emission from a pulsar, whose energy
exceeds the average pulse energy by 10 times or even much more. GPs
were
first detected in the Crab pulsar (PSR B0531+21), which have
nanosecond structure
\citep{1968Sci...162.1481S}. Up to now, this phenomenon has only been
detected in 16 pulsars, including 11 normal pulsars and 5 millisecond
pulsars
\citep{2004IAUS..218..319J,2003ApJ...590L..95J,2001ApJ...557L..93R,
2005ApJ...625..951K,2006AstL...32..583K,2004A&A...427..575K,
2003AstL...29...91E,2006ChJAS...6b..30E}. Generally, the GPs are
related to the high energy emissions (e.g., \citealt{2001ApJ...557L..93R}).
The energy of GPs follows a power-law distribution, while the energy
of normal pulses satisfies a normal or log-normal distribution.
The different pulse energy distribution forms suggest that
the emission mechanisms of them are different \citep{2015ApJ...804L..18R}.

According to the magnetic field strength in the light cylinder $B_{\rm lc}$, the GP-emitting pulsars can be divided into two classes. The first class has
very strong magnetic fields at the light cylinder of $B_{\rm lc}=10^5-10^6\,{\rm G}$
and the second class has relatively weak magnetic fields in the light cylinder
of $B_{\rm lc}=1-100{\rm G}$ \citep{2004A&A...427..575K}.
The first class contains young pulsars (e.g. the Crab pulsar
\citealt{1995ApJ...453..433L}, PSR B0540$-$69
\citealt{2003ApJ...590L..95J}) and millisecond pulsars,
and the second class contains middle-age pulsars
(e.g. PSR B0031$-$07, \citealt{2004A&A...427..575K}).
There is a remarkable difference in GP width between the two classes.
GPs in the first class have short durations with widths on a timescale
of nanoseconds to microseconds, while the typical width of GPs in the
second class is several milliseconds. Theoretically, GPs in the first class can be 
generated by the coherent
instability of plasma near the light
cylinder \citep{2019SCPMA..6279511W}. Some investigators suggest that
GPs in the second group are from the same region as normal pulses \citep{2003AstL...29...91E,2004A&A...427..575K}.

Recently, some pulsars have been reported to exhibit highly unstable behavior in a single pulse. \citet{2015ApJ...804L..18R}
found bright, short-duration radio pulses from
PSR J0901$-$4624 and the energy of the bright pulse follows a power law distribution.
Using the Five-hundred-meter Aperture Spherical radio
Telescope, \citet{2020ApJ...902L..13W} reported that the millisecond pulsar PSR B1534+12
shows two emission states: a weak state and a burst state.
The pulses in the burst state are much narrower and brighter than the weak state.
Also, the pulse energy in the burst state follows a power law distribution, which is different from the weak state.

PSR J1047$-$6709 is a 0.19\,s isolated pulsar with the
$B_{\rm lc}=\rm 7.02\times10^{2}$ G. By
analyzing the pulse energy distribution of
PSR J1047$-$6709, \citet{2012MNRAS.423.1351B}
derived a significant R parameter for this 
pulsar, which is used for giant micropulse 
identification by \citet{2001ApJ...549L.101J}. 
And they suggested that this 
pulsar might be a potential giant micropulse emitter. 
In this paper, we present the first
detection of GPs in this pulsar using the Parkes 64 m radio telescope. 
The observations and data processing are
shown in Section \ref{sec:obs}. In Section
\ref{sec:results}, we present the results. In Section
\ref{sec:discussion}, we discuss and summarise our results.

\section{Observations}\label{sec:obs}

The observational data were downloaded from the Parkes Pulsar Data
Archive that is freely available online
\footnote{\url{https://data.csiro.au}}
\citep{2011PASA...28..202H}. Observations were made on 15 Jan 2012
(MJD 55941), using the Parkes 64-m radio telescope with the center
beam of the Parkes 20-cm Multibeam receiver
\citep{1996PASA...13..243S}
and the fourth-generation Parkes digital filterbank system PDFB4.
The total bandwidth was 256 MHz centred at 1369MHz with
512 channels across the band. A total of 7638
pulses ($\sim$ 1500 seconds) for PSR J1047$-$6709 were obtained.

We used the DSPSR package \citep{2011PASA...28....1V}
to de-disperse using the incoherent dispersion
removal technique and obtain individual-pulse
integrations with the ephemeris from PSRCAT
\citep{2005AJ....129.1993M}.
The ephemeris for PSR J1047$-$6709 were determined by
\cite{1998MNRAS.297...28D}. The RFI were clipped using the paz
plug in of the PSRCHIVE package
\citep{2004PASA...21..302H}. The flux density and polarization
calibration were carried out using the procedure used by
\citet{2011MNRAS.414.2087Y}.

\section{Results}\label{sec:results}
\subsection{The two emission states}

\begin{figure}
\centering
 \includegraphics[width=\columnwidth]{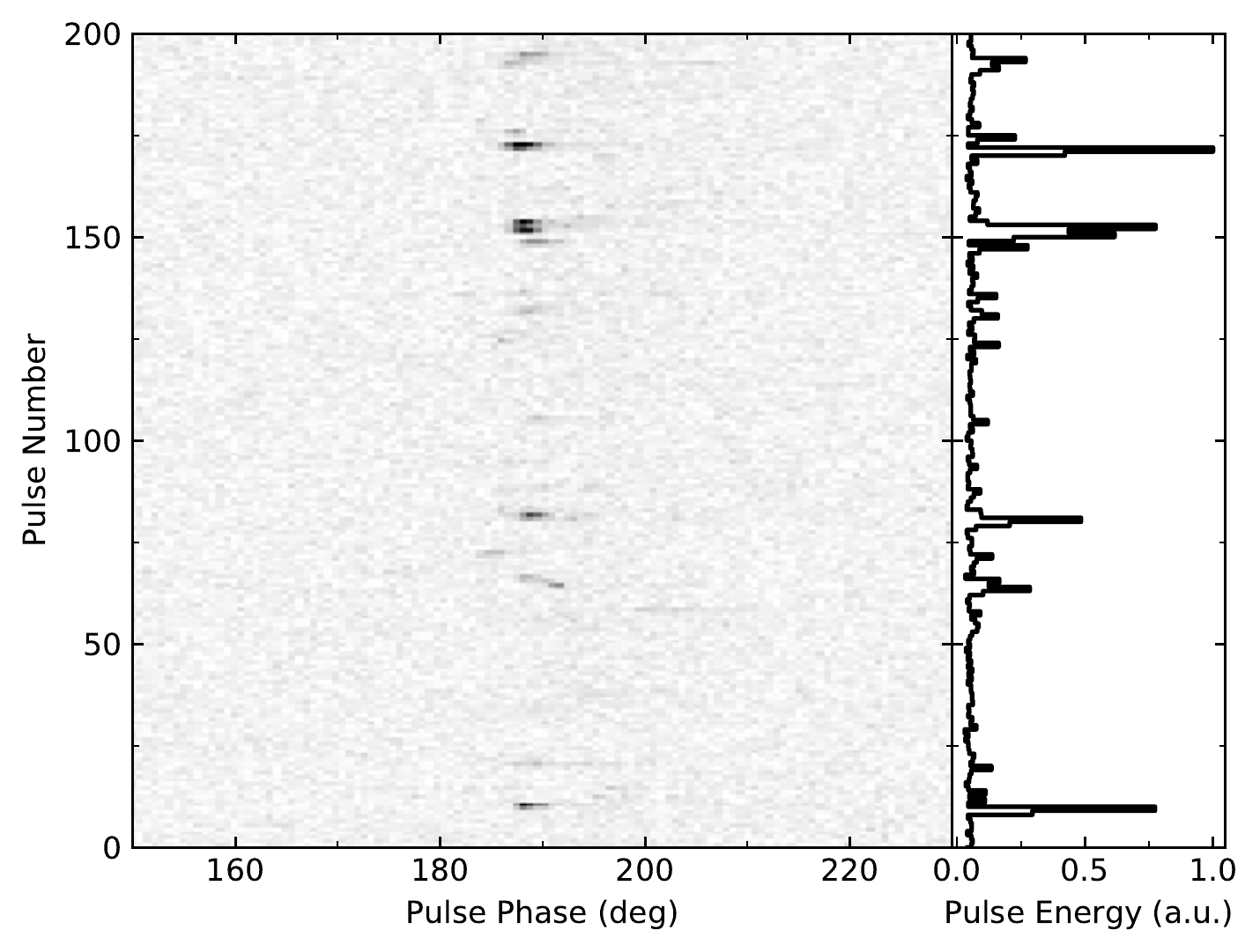}
 \caption{A single-pulse stack of 200 successive pulses for PSR J1047$-$6709.
   The right panel shows the pulse energy variations for the pulse sequence.}
 \label{fig:gray}
 \end{figure}

 \begin{figure}
\centering
 \includegraphics[width=\columnwidth]{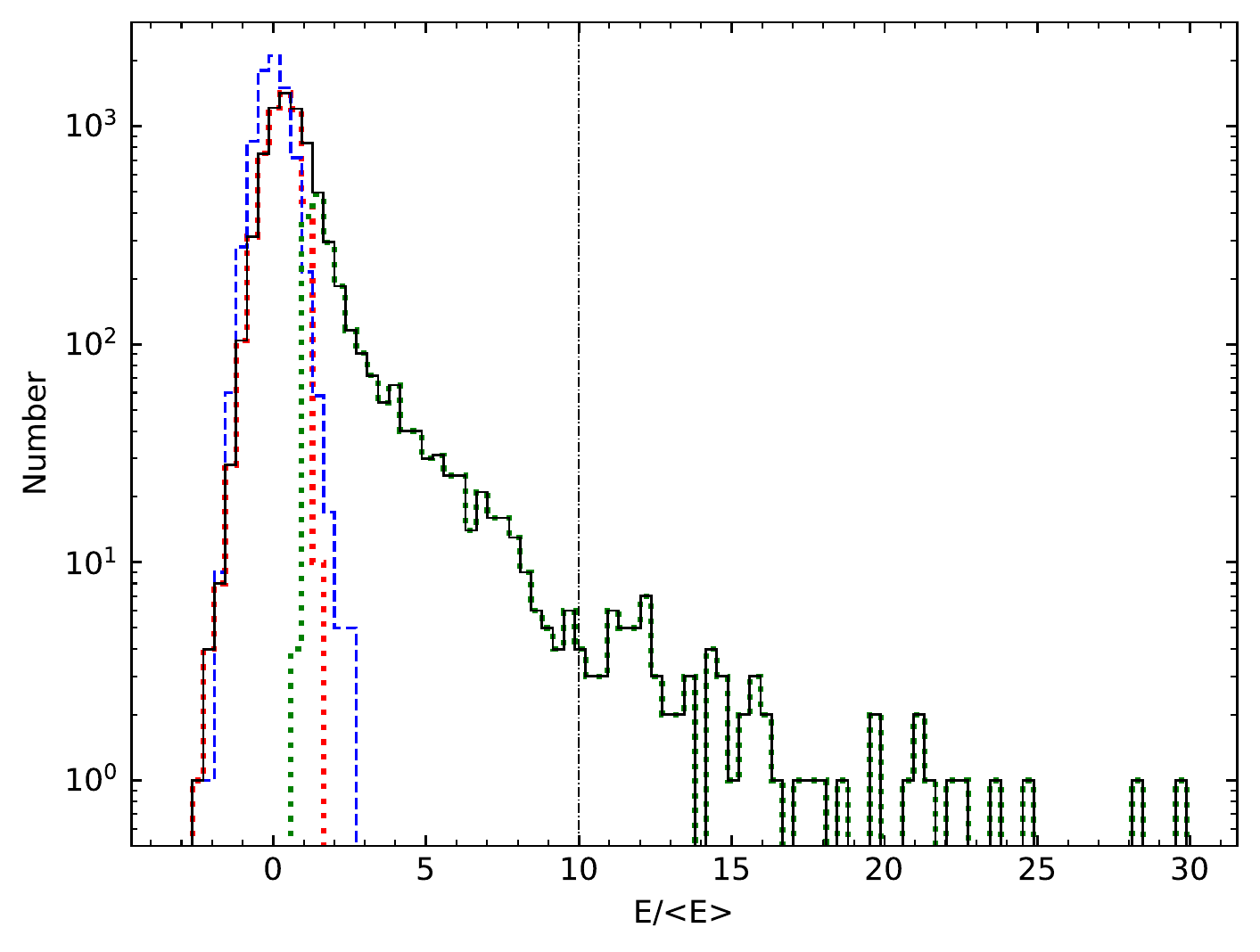}
 \caption{Pulse energy distributions for the off-pulse region (blue dashed histogram),
   the on-pulse region (black solid histogram),
   the weak-state pulses (red dotted histogram)
   and the bright-state pulses (green dotted histogram). In which, the vertical black dot-dashed line represents the 10 times average pulse energy. The energies are normalized by the mean
   on-pulse energy.}
 \label{fig:nullingdist}
 \end{figure}

 A single pulse sequence of 200 individual pulses for PSR J1047$-$6709 is shown in
 Fig. \ref{fig:gray}. The individual pulse energy variations for the same
 pulse sequence are presented in the right panel. The pulse energy
 distributions for the on-pulse and off-pulse windows for all pulses are shown in
 Fig. \ref{fig:nullingdist}. For each individual pulse, the on-pulse energy
 was calculated  by summing the intensities within the on-pulse region after
 subtracting the baseline noise. The off-pulse energy was determined in the
 same way using an equal number of off-pulse bins.
 Fig. \ref{fig:gray} shows dramatic changes in pulse energy, which gives the
 impression that this pulsar might have pulse
 nulling. Generally, the on-pulse energy distribution of a nulling pulsar tends to
 show a peak at zero. However, the peak of on-pulse energy distribution in
 Fig. \ref{fig:nullingdist} is a slightly larger than zero. This implies that the
 apparent null state is probably a weak-emission state instead of a null.

\begin{figure*}
\centering
\includegraphics[width=0.85\textwidth]{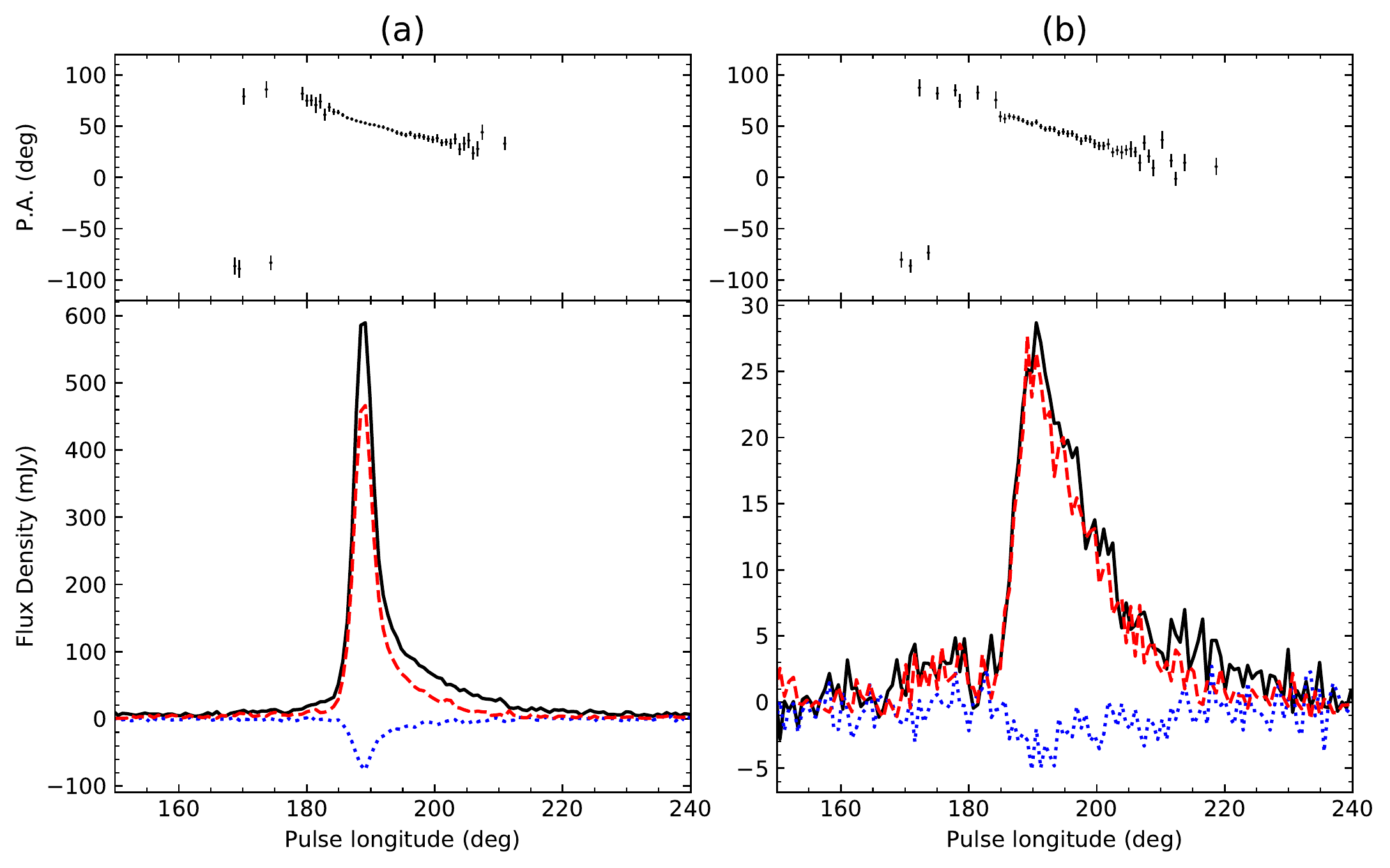}
 \subfigure{
\includegraphics[width=0.85\textwidth]{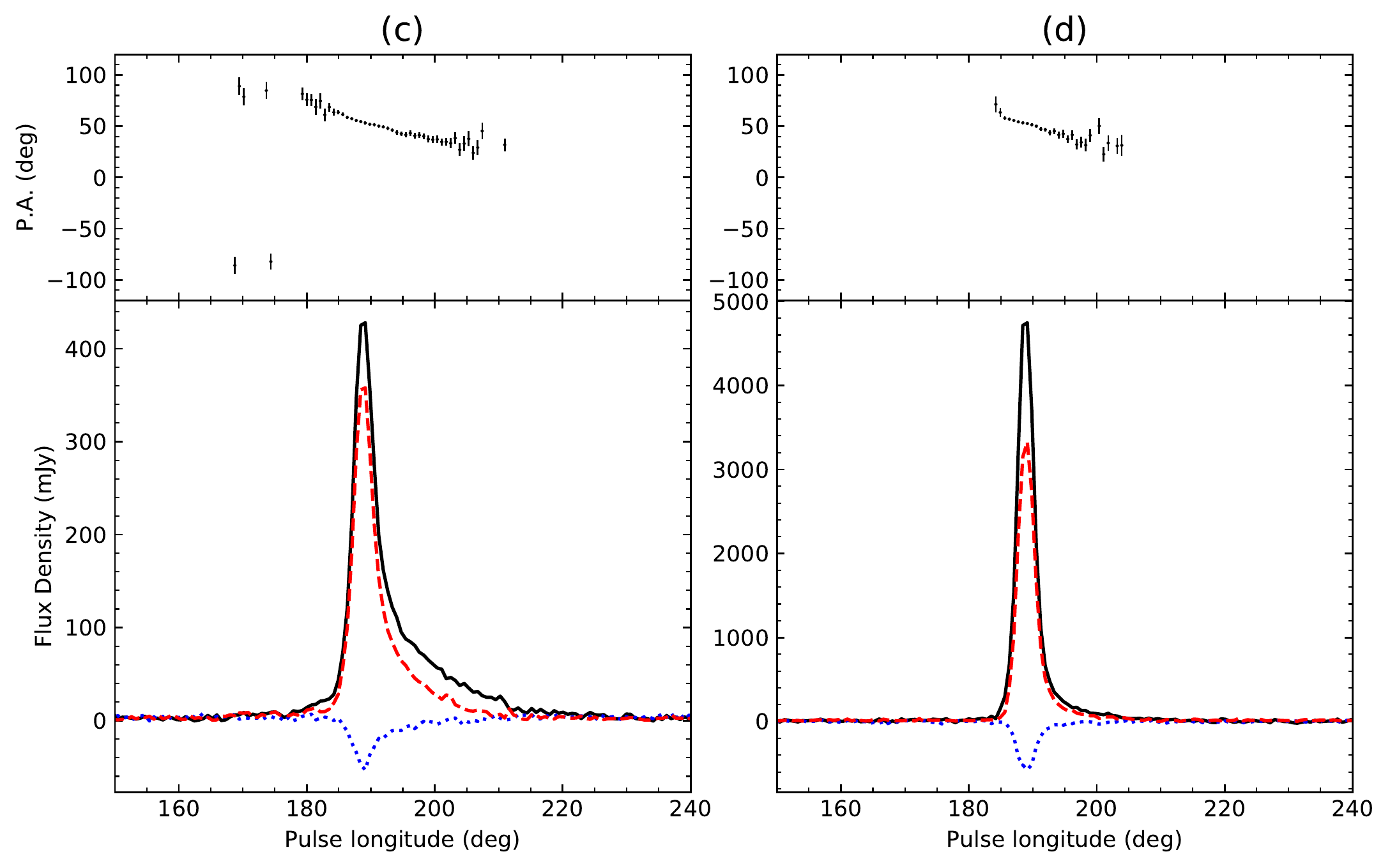}}
\caption{Average polarization properties for (a) the bright state, (b) the weak state, (c) the pulses 
in the bright state with the pulse energy less than 10 times the
  average pulse energy and (d) GPs (pulses in the bright state with the pulse energy larger
  than 10 times the average pulse energy). The lower panels show the pulse profiles for
  total intensity (black line), linearly polarized intensity (red dashed line),
  and circularly polarized intensity (blue dotted line). The upper panels give
  the position angles of the linearly polarized emission.}
 \label{fig:pa}
\end{figure*}

 To further investigate whether the apparent nulls are real nulls or weak states,
 we formed a null-phase pulse profile by average the pulses in the ``null'' state.
 Following \citet{2010MNRAS.408..407B}, we identified the ``null'' pulses
 by comparing the on-pulse energy of individual pulses with the system noise level.
 The uncertainty in the on-pulse energy $\sigma_{\rm on}$ can be expressed as
 $\sqrt{N_{\rm on}}\sigma_{\rm off}$ , where $N_{\rm on}$ is the number of on-pulse
 phase bins, calculated from the mean pulse profile, and $\sigma_{\rm off}$ is the rms
 of the off-pulse region. Pulses with on-pulse energy smaller than
 $3 \sigma_{\rm on}$ were regarded as ``null'' pulses. The average pulse profile obtained
 from all ``null'' pulses is presented in panel (b) of Fig. \ref{fig:pa} which
 shows a clear detection of the pulse profile. We also chose $2 \sigma_{\rm on}$ and
 $1 \sigma_{\rm on}$ as the threshold value to identify ``null'' pulses and the significant
 profile still exists. Therefore, we believe that the apparent null state in
 PSR J1047$-$6709 is a weak-emission state.

\begin{figure}
\centering
\includegraphics[width=\columnwidth]{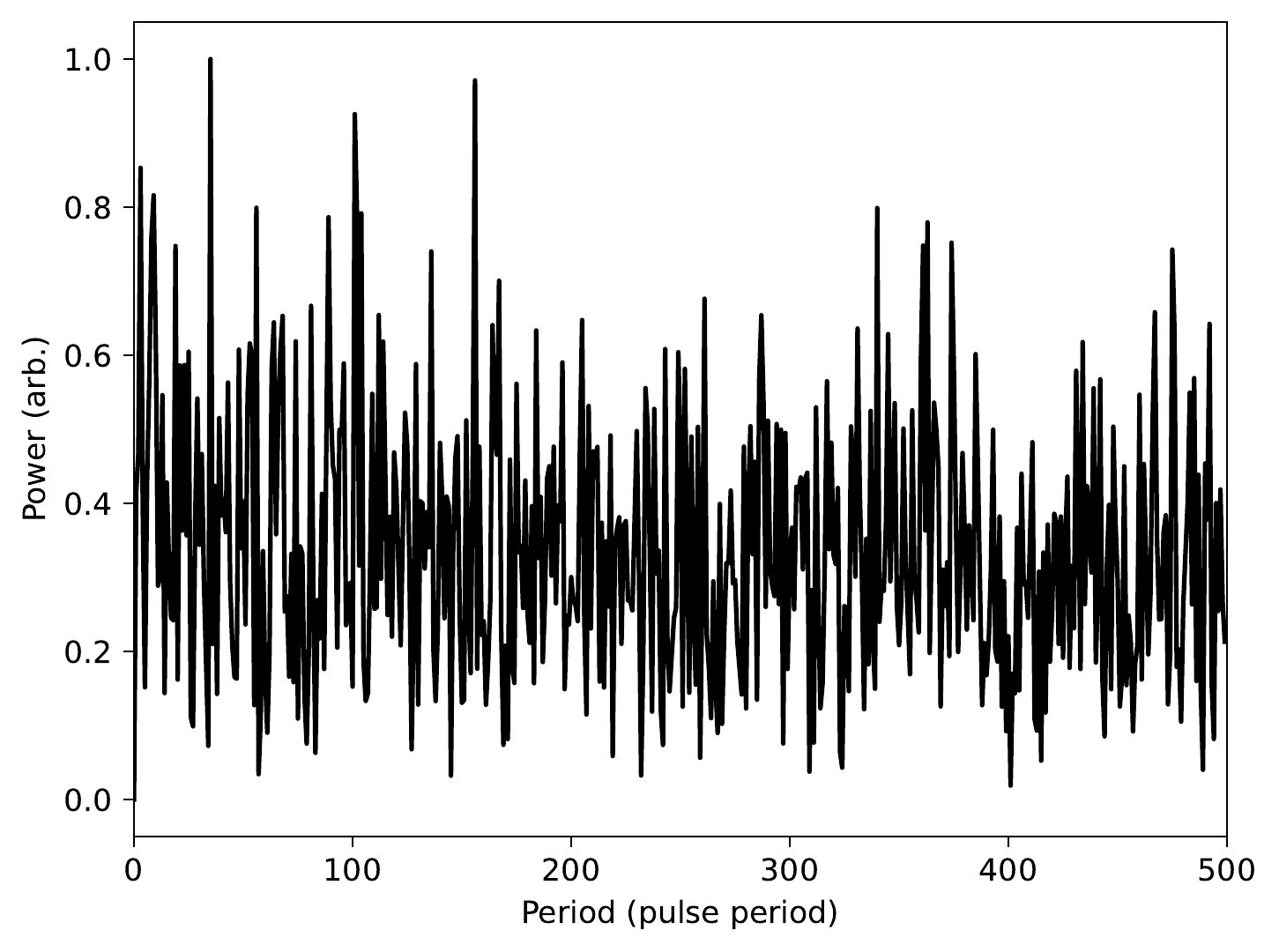}
\caption{Results of fluctuation analysis for PSR J1047$-$6709. 
The spectrum is featureless, without any significant periodicity.}
 \label{fig:fft}
\end{figure}

 Following \citet{2020MNRAS.491.4634Y}, we classified pulses with on-pulse energy smaller than
 $3 \sigma_{\rm on}$  as weak-state pulses and the others were classified as bright-state pulses.
 After the state separation, we found that this pulsar spends 28 per cent of the time in the bright state and 72 per cent of the time in the weak state
 during our observation. The pulse energy distributions of the two emission states are shown
 in Fig. \ref{fig:nullingdist}. The pulse energy of the weak state shows
 a Gaussian distribution while the pulse energy distribution
 of the bright state can be described by a power-law distribution.
 The different pulse energy distributions suggest that the emission mechanisms for the two
 states are different. 
 
We used the RMFIT program of 
PSRCHIVE to estimate the RM value of this pulsar. Our result is $-78.7\pm0.1 \rm \ rad \ m^{-2}$ which
is more accurate than the psrcat value $-79.3\pm2 \rm \ rad \ m^{-2}$ \citep{2008MNRAS.386.1881N}. Therefore, we used $-78.7 \rm \ rad\ m^{-2}$
to correct the Faraday rotation effect. 
 Average pulse profiles for both states are shown in
 Fig. \ref{fig:pa}. The pulse profile for the bright state is significantly narrower than that
 for the weak state. The half-power pulse widths (W50) are $4\fdg5$ (2.5 ms) and $13\fdg2$
 (7.3 ms), respectively. The  peak flux density of the weak state is 28.7 mJy which is 
 about one twentieth of the peak flux density of the bright state. The fractional linear 
 polarization for the weak state is nearly 100\% which is larger than that for the 
 bright state (68\%). We calculated the absolute circular polarization fraction by 
 the ratio of the mean absolute circular polarization intensity 
 $\langle\lvert V \lvert\rangle$ and the mean flux density S. We obtained the fractional absolute circular polarization for the weak state is 17.27\%, which is somewhat larger than that of the bright state  of 11.42\%. The differences in linear and circular polarization could be an indication of different emission mechanism for the two states. However, the position angles (PAs) of the linear polarization for the 
 two states show similar variations with pulse phase. This suggests that the emission 
 geometry is basically the same for the two emission states.

We searched for periodicity in the occurrence time of the pulse in the bright state for PSR J1047$-$6709. We carried out an analysis of fluctuation spectra based on the discrete Fourier transform (DFT), which can be used to determine the modulation period. The results of the DFT are shown in Fig. \ref{fig:fft}, in which there is no significant peaks are detected. Therefore, the occurrence time of the pulse in the bright state is non-periodical.

 \subsection{Giant pulse}
In recent years, some researchers find that GPs are characterized by their high brightness temperature, short duration time and power-low distribution \citep{2006ChJAS...6b..41K,2015ApJ...803...83B,2019MNRAS.483.4784M}. More traditionally, GPs can also be defined
as very bright single pulses with the energy larger than 10 times the average
pulse energy \citep{2012AJ....144..155S}.
The time resolution of our data was not sufficient to resolve the short timescale structure, so we use the traditional definition of GPs here for our analysis.
\citet{2012MNRAS.423.1351B} reported that PSR J1047$-$6709 may be a potential giant 
micropulse emitter.
Fig. \ref{fig:nullingdist} shows that the single pulses have pulse energy exceeding 10 times
the average pulse energy and those pulses can be defined as GPs (the 10 times
the average pulse energy is labeled by the  vertical black dot-dashed 
line in the Fig. \ref{fig:nullingdist}). A total
of 75 GPs were detected in our observations. This is the first report that GPs
are detected in PSR J1047$-$6709. We also noted that the scintillation bandwidth for PSR J1047$-$6709
estimated by the NE2001 model \citep{2002astro.ph..7156C,2003astro.ph..1598C} at the
frequency of 1 GHz is about 300 KHz, far less than the observing bandwidth, so the scintillation
effect can be ignored here.

\begin{figure}
\centering
 \includegraphics[width=\columnwidth]{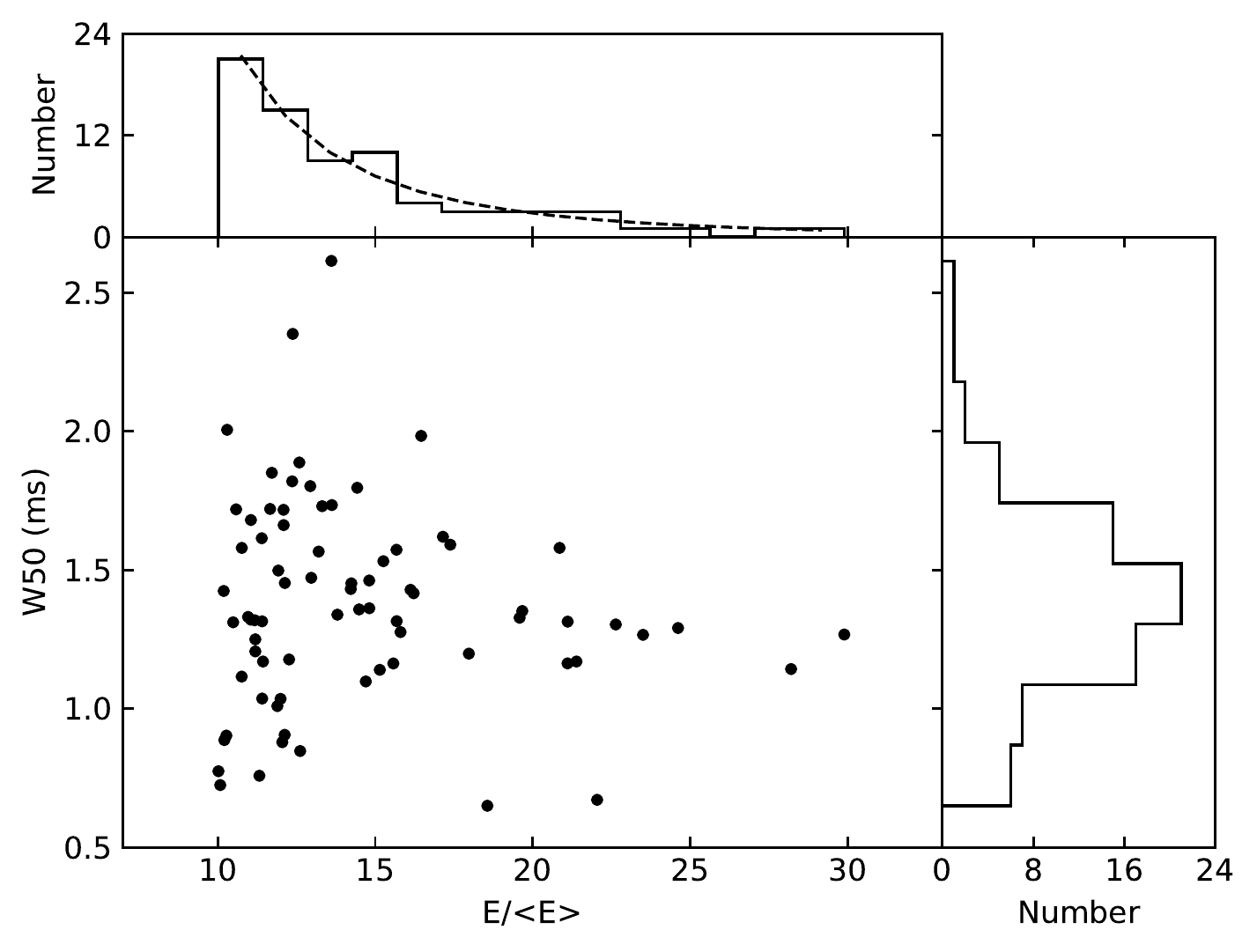}
 \caption{ The relative energy versus the W50 of 75 
 GPs. The upper panel: the histograms of the energy 
 which has an approximate power-law distribution with 
 a index of $\alpha=-3.26\pm0.22$ by a dashed line; 
 The right panel: the W50 histograms of the GPs.}
 \label{fig:ew}
 \end{figure}

Fig. \ref{fig:ew} shows the distributions of pulse width and energy for all 75 GPs.
The GPs are relatively narrow, most of which have W50 smaller than 2 ms. By comparison, W50 of the
average profile for all pulses is 2.9 ms. As shown in the top panel of Fig. \ref{fig:ew},
the pulse energy distribution for the GPs can be well
fitted by a power-law distribution with the index $\alpha=-3.26\pm0.22$. This index agrees well
with the Crab pulsar. The peak flux density of these GPs are in the range of 17 to 110 times brighter
than that of the mean pulse profile. The peak flux density of the brightest GP is 19 Jy which is 110 times stronger than the mean pulse profile.

The arrival time of GPs can usually be described using a Poisson distribution
\citep{1995ApJ...453..433L,2007MNRAS.378..723K}. The probability of obtaining GPs in a
Poisson distribution is:
\begin{equation}\label{1}
  P(K)=\frac{e^{-\lambda}\lambda^{K}}{K!},
\end{equation}
where $\lambda$ is the interval of the expected GPs number. And K is the number of GPs
observed over an interval of time. $\chi^{2}$ statistic can be used as a fitting
optimization index:
\begin{equation}\label{2}
 \chi^{2}= \sum_{k}\frac{(N_{k}-n_{k})^{2}}{n_{k}},
\end{equation}
where k is the iteration of different GPs numbers occurring in an interval. $n_{k}$
is the frequency of GPs in each count predicted by poisson model, which can
be calculated from  $n_{k}=P(K)\sum_{i}N_{i}$, $N_{k}$ is the observed frequency of
GPs in each count.

To distinguish whether the arrival time of the GPs from PSR J1047$-$6709 can be described by a Poisson distribution, the 1500\,s pulse sequence was analyzed. We binned our data with bin of width 25\,s and the results are shown in Table \ref{tabel:1}.
The $\chi^{2}$ value was 7.95 with the confidence level of 90\%. We also analyzed our data with bins of width 50\,s and 75 s and the $\chi^{2}$ value and confidence levels of them were 6.59, 12.20 and 50\%, 75\%, respectively. The arrival time of the GPs from this pulsar possibly follows a Poisson distribution. More observations are needed to confirm this possibility.

We also compared the polarization properties of the GPs to those pulses in the
bright state with the pulse energy less than 10 times the average pulse energy.
The average polarization profiles for these two pulse classes are
shown in the panels (c) and (d) of Fig. \ref{fig:pa}, respectively. Although the GP profile is relatively
narrow, they have similar profile shapes. The linear polarization intensities for both
of them are high. Their PA swings and circular polarization are also similar. These similarities
suggest that the emission mechanism is basically the same for GPs and the pulses 
in bright state with enengy less than 10 times average pulse enenrgy, which supports the idea that 
GPs are generated in the polar gap region for this pulsar.

\begin {table}
\caption{Arrival time statistics of all 1500 s time series with bin
of width 25 s. Column (1) is the number of GPs detected in an interval (K). 
Column (2) is the number of times observed in an interval with K GPs. 
Column (3) is the number of times predicted by Poisson distribution 
formula (\ref{1}).}
\label{t4}
\begin{tabular}{ccc}
  \hline
  \multicolumn{3}{c}{$\rm t_{bin}=25 s $} \\
  \hline
  \thead[c]{Pulses in interval \\ (k) \\ (1)} &  \thead[c]{ Observed frequency \\ ($ N_{k})$ \\ (2)}
   &\thead[c]{Poisson prediction \\ ($ n_{k})$ \\ (3)} \\
   \hline
   0 & 21 &  17.19   \\
   1 & 18 &  21.48   \\
   2 & 11 &  13.42   \\
   3 & 5  &  5.59    \\
   4 & 5  &  1.74    \\
   5 & 0  &  0.0     \\
   6 & 0  &  0.0     \\
   7 & 0  &  0.0     \\
   8 & 0  &  0.0     \\
   9 & 0  &  0.0     \\
  \hline
\end{tabular}
\label{tabel:1}
\end{table}

\section{Discussion and conclusions}\label{sec:discussion}

We reported the two emission states for PSR J1047$-$6709 for the first time. This pulsar switches
between a weak state and a bright state.
The pulse energy in the bright state follows a power-law distribution while that in the weak state
follows a Gaussian distribution. The different pulse energy distributions suggest that the emission
physics of the two emission states are different. The polarization properties of the two states are
also different.
However, the PA swings for the two states are similar. This
implies that the magnetospheric field geometry for this pulsar is unchanged, even if the emission
switches between two different states. We argue that state switches for PSR J1047$-$6709 are
related to the variations of the current in the magnetospheric field, not the magnetospheric
field geometry.

During the bright state of PSR J1047$-$6709, we identified 75 GPs whose energies are 10 times
larger than the average pulse energy. The peak flux density of the brightest GP is about 19 Jy which
is 110 times higher than the peak flux density of the mean pulse profile. We compared these
GPs to the pulses in bright state with enengy less than 10 times average pulse enenrgy,
and found that these two classes may have a same physical origin. The GPs of
PSR J1047$-$6709 are only a tail of the power-law energy distribution of the pulses in
the bright state. One possibility is that the traditional definition of GPs which is only based on the pulse energy is not suitable for this pulsar, and the GPs we detected are actually normal pulses instead of real ``giant" pulses. Many authors also realize that pluse energy is not a good criterion for distinguishing GPs \citep{2006ChJAS...6b..41K,2015ApJ...803...83B,2019MNRAS.483.4784M}. High time-resolution observations are needed to clarify the nature of GPs we detected in this paper.

As mentioned in Section \ref{sec:intro}, GP pulsars can be divided into two classes according to the magnetic field in the light cylinder ($B_{\rm lc}$). PSR J1047$-$6709 belongs
to the class of middle-aged pulsars. 
The GP widths of this pulsar are in the range of several hundred microseconds to several milliseconds, which is similar to the typical GP width of middle-aged GP pulsars~\citep{2004A&A...427..575K}, but much wider than that of the Crab pulsar or millisecond pulsar~\citep{1968Sci...162.1481S, 2019MNRAS.483.4784M}.  
Generally, the $B_{\rm lc}$ for the middle-aged GP pulsars is weak, in the range of 1 to 100 G. The $B_{\rm lc}$ for PSR J1047$-$6709 is about 700\,G which is stronger than that of other middle-aged GP pulsars.
Also, this pulsar have the shortest spin period in the middle-aged GP pulsar class.

The evolutionary relation for the two GP-emitting pulsar classes is still unclear. According to the typical pulsar evolution model,
a young pulsar (such as the Crab pulsar) will evolve to a middle age pulsar~\citep{2012puas.book.....L}. If there
are some relations between the two classes, the nanosecond GPs which were typically
detected in the first class would possibly be detected in the second class. However,
the nanosecond GPs have only been seen in the first class so far. Maybe the magnetic
environment that can generate the nanosecond GPs breaks with the pulsar evolution.
Alternatively, the two classes may do not have an evolutionary relation. At present,
the GP-emitting pulsars are still rare and more samples are needed to expose their
intrinsic physics.

\section*{Acknowledgments}
We would like to take this opportunity to thank an anonymous referee for all his useful comments. We appreciate Feifei Kou useful discussion. The Parkes radio telescope is part of the Australia Telescope National Facility which is funded
by the Commonwealth of Australia for operation as a National Facility managed by CSIRO. This work
is supported by the 2021 open program of the Key Laboratory of Xinjiang Uygur Autonomous Region, the National Natural Science Foundation of China (No. U1831102, U1731238, U1838109, 11873080), the CAS “Light of West China” Program (No. 2017-XBQNXZ-B-022, 2018-XBQNXZ-B-023,) and the Strategic Priority Research Program of Chinese Academy of Sciences (No. XDB23010200).

%\bibliographystyle{mnras}
%\bibliography{ref,journals}

\section*{DATA AVAILABILITY}
The data underlying this article will be shared on reasonable request
to the corresponding author.

\bsp	
\label{lastpage}
\end{document}